      [1999/12/01 v1.4c Il Nuovo Cimento]
\documentclass[final]{cimento}

\usepackage{graphicx}  
\title{Particle acceleration, magnetic field generation,
and emission in relativistic pair jets}
\author{K.-I.~Nishikawa\from{ins:x}\ETC,
P.~Hardee\from{ins:y}, C. B.~Hededal\from{ins:z},
G.~Richardson\from{ins:a}\\
H.~Sol\from{ins:b}, R.~Preece\from{ins:c}
        \atque
G. J.~Fishman\from{ins:d}}  
\instlist{\inst{ins:x} National Space Science and Technology Center,
  Huntsville, AL 35805 USA
\inst{ins:y} Department of Physics and Astronomy,
  The University of Alabama,
  Tuscaloosa, AL 35487 USA
\inst{ins:z} Niels Bohr Institute, Department of Astrophysics,
Juliane Maries Vej30, 2100 K\o benhavn \O, Denmark

\inst{ins:a} Department of Mechanical and Aerospace Engineering
  University of Alabama in Huntsville Huntsville, AL 35899 USA

\inst{ins:b} LUTH, Observatore de Paris-Meudon, 5 place Jules Jansen
  92195 Meudon Cedex, France

\inst{ins:c} Department of Physics,
  University of Alabama in Huntsville,
  Huntsville, AL 35899 and National Space Science and Technology Center,
  Huntsville, AL 35805 USA

\inst{ins:d} NASA-Marshall Space Flight Center, \\
National Space Science and Technology Center,
  Huntsville, AL 35805 USA

inst{ins:y}} \PACSes{\PACSit{00.00}{By the way, which PACS is it,
the 00.00? GOK.} \PACSit{---.---}{\ldots}}
\begin{document}

\maketitle

\begin{abstract}
Shock acceleration is a ubiquitous phenomenon in astrophysical
plasmas.  Plasma waves and their associated instabilities (e.g.,
Buneman, Weibel and other two-stream instabilities) created in
collisionless shocks are responsible for particle (electron,
positron, and ion) acceleration. Using a 3-D relativistic
electromagnetic particle (REMP) code, we have investigated particle
acceleration associated with a relativistic  jet front propagating
into an ambient plasma. We find that the growth times of Weibel
instability are proportional to the Lorentz factors of jets.
Simulations show that
the Weibel instability created in the collisionless shock front
accelerates jet and ambient particles both perpendicular and
parallel to the jet propagation direction. The small scale magnetic
field structure generated by the Weibel instability is appropriate
to the generation of ``jitter'' radiation from deflected electrons
(positrons) as opposed to synchrotron radiation. The jitter
radiation resulting from small scale magnetic field structures may
be important for understanding the complex time structure and
spectral evolution observed in gamma-ray bursts or other
astrophysical sources containing relativistic jets and relativistic
collisionless shocks.
\end{abstract}

\section{Introduction}
Nonthermal radiation observed from astrophysical systems containing
relativistic jets and shocks, e.g., active galactic nuclei (AGNs),
gamma-ray bursts (GRBs), and Galactic microquasar systems usually
has power-law emission spectra. In most of these systems, the
emission is thought to be generated by accelerated electrons through
the synchrotron and/or inverse Compton mechanisms. Radiation from
these systems is observed in the radio through the gamma-ray region.
Radiation in optical and higher frequencies typically requires
particle re-acceleration in order to counter radiative losses.

Particle-in-cell (PIC) simulations can shed light on the physical
mechanism of particle acceleration that occurs in the complicated
dynamics within relativistic shocks.  Recent PIC simulations using
injected relativistic electron-ion jets show that acceleration
occurs within the downstream jet, rather than by the scattering of
particles back and forth across the shock as in Fermi acceleration
[1-7]. In general, these independent simulations have confirmed that
relativistic jets excite the Weibel instability [8]. The Weibel
instability generates current filaments with associated magnetic
fields [9], and accelerates electrons [1-7].

In this paper we present new simulation results of particle
acceleration and magnetic field generation for relativistic
electron-positron shocks using 3-D relativistic electromagnetic
particle-in-cell (REMP) simulations. In our new simulations, the
growthrates with different Lorentz factors of jets have been studied
without an initial ambient magnetic field.

\section{Simulation Setup and results}

Two simulations were performed using an $85 \times 85 \times 320$
grid with a total of 180 million particles (27
particles$/$cell$/$species for the ambient plasma) and an electron
skin depth, $\lambda_{\rm ce} = c/\omega_{\rm pe} = 9.6\Delta$,
where $\omega_{\rm pe} = (4\pi e^{2}n_{\rm e}/m_{\rm e})^{1/2}$ is
the electron plasma frequency and $\Delta$ is the grid size
(Nishikawa et al. 2004).
%

The electron number density of the jet is $0.741n_{\rm b}$, where
$n_{\rm b}$ is the density of ambient (background) electrons. The
average jet velocities for the two simulations are $v_{\rm j} =
0.9798c, 0.9977c$ corresponding to  Lorentz factors are 5 (2.5 MeV)
and 15 (7.5 MeV), respectively. The jets are cold ($v^{\rm e}_{\rm
j, th} = v^{\rm p}_{\rm j, th} = 0.01c$ and $v^{\rm i}_{\rm j, th} =
0.0022c$) in the rest frame of the ambient plasma. Electron-positron
plasmas have mass ratio $m_{\rm p}/m_{\rm e} = 1$. The thermal
velocity in the ambient plasmas is $v^{\rm e, p}_{\rm th} = 0.1c$
where $c$ is the speed of light. The time step $\Delta t =
0.013/\omega_{\rm pe}$.


Current filaments resulting from development of the Weibel
instability behind the jet front are shown in Figs. 1a  and 1b at
time $t = 28.8/\omega_{\rm pe}$ for unmagnetized ambient plasmas (a)
$\gamma = 5$ and (b) $\gamma = 15$. The maximum values of $J_{\rm
y}$ are (a) 15.63 and (b) 3.86, respectively. The slower jet shows
larger amplitudes than the faster jet at the same simulation time.
The effect of Lorentz factors of jets affects the growth rates of
Weibel instability as expected by the theory [8].

The differences in the growthrates between the differerent Lorentz
factors are seen more clearly in the $x$-component of the generated
magnetic fields as shown in Fig. 2. The amplitudes of $B_{\rm x}$ in
the slower jet (a) are much larger than those in the faster jet (b).
The comparisons in the saturated magnetic fields need much longer
simulations, which are in progress at the present time.


\begin{figure}
\hspace*{1.2cm}
\includegraphics[height=.52\textheight]{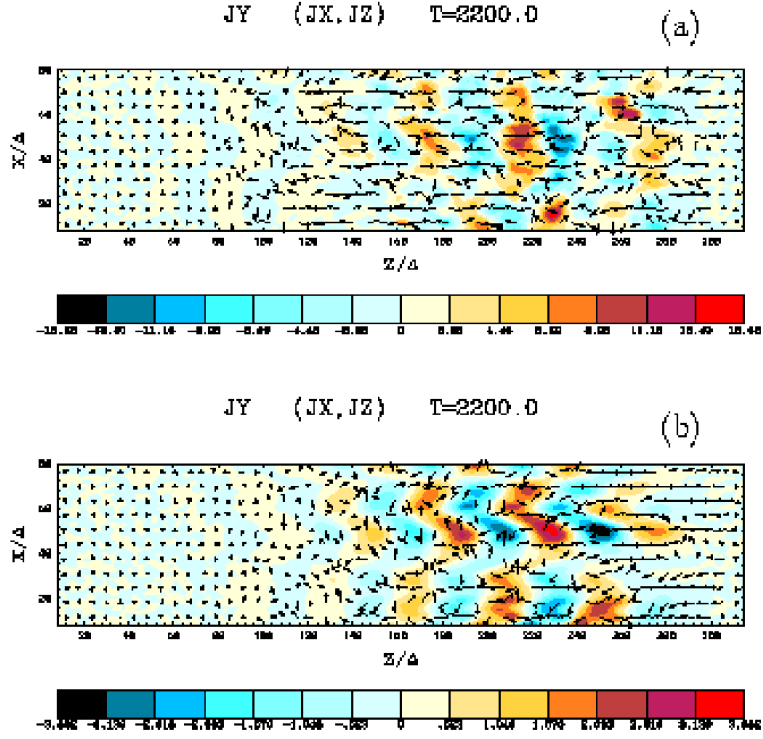}     
\caption{2D images show the current density ($J_{\rm y}$) at $t =
28.8/\omega_{\rm pe}$ (a) $\gamma = 5$ and (b) $\gamma = 15$. Colors
indicate the $y$-component of the current density, $J_{\rm y}$
[peak: (a) 15.63 and (b) 3.86], and the arrows indicate $J_{\rm z}$
and $J_{\rm x}$.}
\end{figure}

\begin{figure}
\includegraphics[height=.205\textheight]{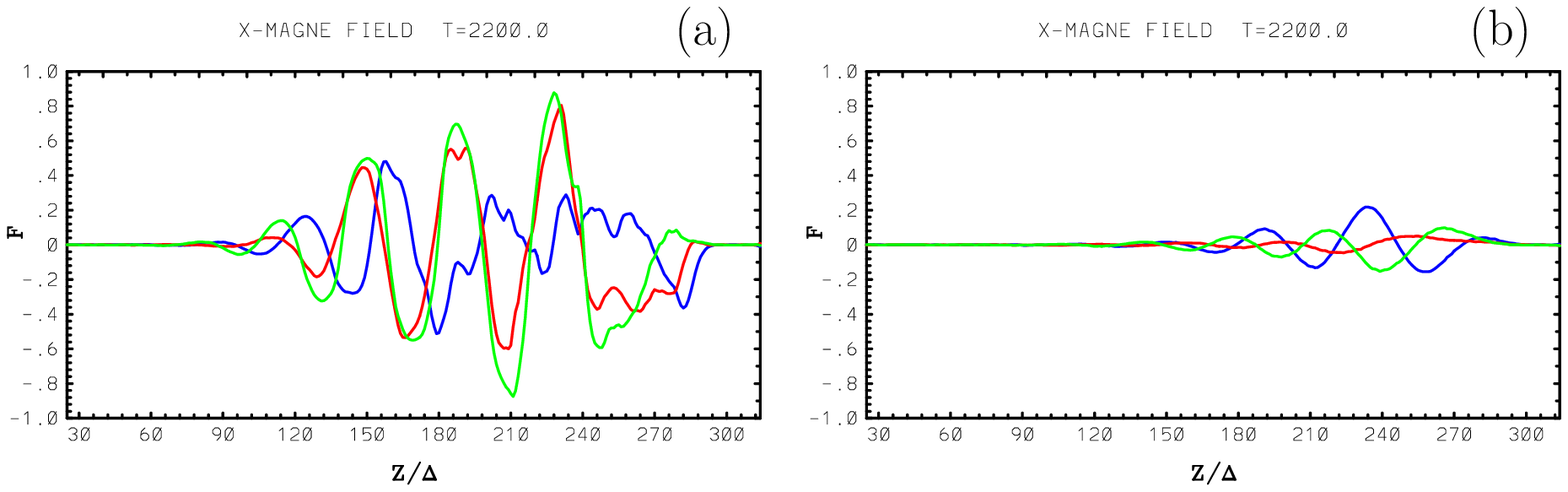}     
\caption{One-dimensional cuts along the $z$-direction ($25 \leq
z/\Delta \leq 314$) of a flat jet (a) $\gamma = 5$ and (b) $\gamma =
15$. Shown are the $x$-component of the magnetic field shown at $t =
28.8/\omega_{\rm pe}$. Cuts are taken at $x/\Delta = 38$ and
$y/\Delta = 33$ (blue-dotted), 43 (red-solid), 53 (green-dashed) and
separated by about one electron skin depth.}
\end{figure}


\section{Summary and Discussion}
We have performed self-consistent, three-dimensional relativistic
particle simulations of relativistic electron-positron jets
propagating into magnetized and unmagnetized electron-positron
ambient plasmas. The main acceleration of electrons takes place in
the region behind the shock front [5]. Processes in the relativistic
collisionless shock are dominated by structures produced by the
Weibel instability.  This instability is excited in the downstream
region behind the jet head, where electron density perturbations
lead to the formation of current filaments. The nonuniform electric
field and magnetic field structures associated with these current
filaments decelerate the jet electrons and positrons, while
accelerating the ambient electrons and positrons, and accelerating
(heating) the jet and ambient electrons and positrons in the
transverse direction.

The growthrates depend on the Lorentz factors of jet as expected by
the theory [9].  The $e$-fold time is written as $\tau \simeq
\sqrt{\gamma_{\rm sh}}/\omega_{\rm pe}$, where $\gamma_{\rm sh}$ is
the Lorentz factor of the shock. The simulation results show that
$\tau \propto \gamma_{\rm sh}$.

Other simulations with different skin depths and plasma frequencies
confirm that both simulations have enough resolutions and the
electron Weibel instability is characterlized by the electron skin
depth [5].

These simulation studies have provided new insights for particle
acceleration and magnetic field generation. Further research is
required to develop radiation models based on these microscopic
processes.

\acknowledgments K. Nishikawa is a NRC Senior Research Fellow at
NASA Marshall Space Flight Center. This research (K.N.) is partially
supported by the National Science Foundation awards ATM-0100997, and
INT-9981508. P. Hardee acknowledges partial support by a National
Space Science and Technology (NSSTC/NASA) award.  The simulations
have been performed on IBM p690 (Copper) at the National Center for
Supercomputing Applications (NCSA) which is supported by the
National Science Foundation.

\end{document}